\documentclass[12pt]{article}
\title {Bounce solutions and the transition to thermal hopping in
$\varphi^4$ theory with a $\varphi^3$ term}
\author{Hatem Widyan\thanks{E--mail : hatem@ducos.ernet.in} ,
A. Mukherjee\thanks{E--mail : am@ducos.ernet.in} ,
N. Panchapakesan\thanks{E--mail : panchu@ducos.ernet.in} and
R. P. Saxena\thanks{E--mail : rps@ducos.ernet.in}  \\
	{\em Department of Physics and Astrophysics,} \\
	{\em University of Delhi, Delhi-110 007, India} 
	}

\begin {document}
\maketitle
\begin {abstract}
The nature of the transition from quantum tunneling at low
temperatures to thermal hopping at high temperatures is investigated
in a scalar field theory with cubic symmetry breaking. The bounce
solution which interpolates between the zero-temperature and
high-temperature solutions is obtained numerically, using a multigrid
method. It is found that, for a small value of the symmetry-breaking
coupling $f$, the transition is first-order. For higher values of $f$,
the transition continues to be first-order, though weakly so.
\end{abstract}
\begin {section}*{I. INTRODUCTION}

\par
The problem of decay of a metastable state via quantum tunneling 
has important 
applications in many branches of physics, from condensed matter to particle 
physics and cosmology.
In the semi-classical approximation, the decay rate per unit 
volume is given by an expression of the form
\begin{equation}
  \Gamma \quad = \quad A \quad e^{-S_E} , \label{eq:decay}
\end{equation}
where $S_E$ is the Euclidean action of the bounce: 
the classical solution of the equation of motion with appropriate boundary 
conditions. The bounce has turning points at the configurations at which 
the system enters and exits the potential barrier, and the analytic 
continuation to Lorentzian time at the exit point gives us the configuration 
of the system at that point and its subsequent evolution.  
The solution of the 
equation of motion looks like a bubble in four dimensional Euclidean space 
with radius R and thickness proportional to the parameter in the symmetry 
breaking term. If there is more than one solution satisfying the 
boundary conditions, the one with the lowest $S_E$
will dominate Eq.~(\ref{eq:decay}). 
 The prefactor A comes from Gaussian functional integration over small 
fluctuations around the bounce. The formalism at 
zero temperature is well known \cite{Langer,Coleman}; the least action is 
given by the bounce which is O(4) invariant  \cite{Glasser}.

 At nonzero temperature, the above formalism has to be modified, since
the metastable state can decay due to classical thermal motion. In the
context of quantum mechanics, the extension of the formalism to
finite temperature was carried out by Affleck \cite{Affleck} . 
He argued that
a second order phase transition from the quantum to the thermal regime takes
place at some critical temperature. Chudnovsky \cite{Chudnovsky}
showed that this is not always true. Depending on the shape of 
the potential barrier, the crossover from thermally assisted quantum 
tunneling to thermal activation at high
temperature is either first-order
(i.e., the first derivative of $\Gamma$ at the transition 
temperature is discontinuous) or second-order (i.e., the first 
derivative of $\Gamma$ is continuous, but the second derivative 
is discontinuous at the transition temperature).

 In the context of quantum field theory, little work has been done 
to explore the nature of the transition from the quantum regime 
(zero-temperature) to the thermal regime (high-temperature). Linde 
\cite{Linde} suggested that periodic 
bounces could \emph{smoothly} interpolate between these two regimes. In this 
picture, at zero temperature the solution is an O(4) symmetric bubble 
with a  
radius R. Up to $T \sim {(2R)}^{-1}$, we have periodic, widely separated 
bounces, and beyond this temperature they start merging into one another 
producing what is known as a ``wiggly cylinder'' solution. As one keeps 
increasing the temperature these wiggles smoothly straighten out and the 
solution goes into an O(3) invariant cylinder (independent of Euclidean time 
$\tau$), which dominates the thermal activation regime.

 Garriga \cite{Garriga} extended the work of Chudnovsky by studying
bubbles in the thin-wall approximation (TWA) at high and low
temperatures, as well as the wiggly cylinder solution at intermediate
temperatures. He showed that in the TWA the transition is always
first-order. However, although motivated by quantum field theory, 
this work wan not truly field-theoretical.

 Ferrera \cite{Ferrera} carried the investigation beyond the domain of
validity of the TWA. In fact, this was the first fully field-theoretic
investigation of bubble formation at arbitrary temperatures. (Such a
study must solve the equation of motion at intermediate temperatures,
which requires solving a partial differential equation.) Ferrera
studied a model field theory which was $\varphi^4$ theory with a
symmetry-breaking term of the form $f\varphi$. His result is that 
only for very large wall thickness (i.e., large $f$)
a second-order phase transition takes place, while for all other 
cases a first-order phase transition occurs.

 In an earlier paper \cite{Hatem}, we studied phase transitions in
two kinds of $\varphi^4$ theory, with symmetry-breaking terms proportional
to $\varphi$ and $\varphi^3$ respectively. We obtained accurate 
numerical solutions to the
equation of motion in the zero- and high-temperature limits, and
found that, even for a fairly large value of the $\varphi^3$ coupling,
the thin-wall approximation (TWA) holds to a few percent. An analytical
solution for the bounce was obtained, which reproduces the action in
the thin-wall as well as the thick-wall limits. We also
investigated the dependence of $T_\star$, defined by
$S_4/S_3=1/T_\star$, on the dimensionless symmetry breaking parameter $f$.
 For small $f$, the TWA behavior $T_\star \propto f$ is obtained, while at
larger values of $f$ there is a smooth departure from this \cite{Hatem}.

 It has been argued recently that the leading correction 
to the tree potential due to one-loop effects at finite temperature
is proportional to $\varphi^3$ rather than $\varphi$ \cite{Dine}. It therefore
becomes natural to investigate $\varphi^4$ theory with $\varphi^3$
symmetry-breaking at arbitrary temperatures. In the present paper, we
undertake such an investigation. Our aim is to obtain the bounce
solutions at all temperatures, to compute the action and to determine
the nature of the phase transition from quantum tunneling to thermal
hopping. Our work therefore follows a parallel track to that of 
Ferrera \cite{Ferrera}, and
it is of interest to compare the results for the two theories. We find
that, for small values of $f$, the action shows a kink when plotted
against temperature or its inverse $\beta$. This can be identified as
a first-order transition. For larger values of $f$ the transition is
weakly first-order. The second-order behavior seen by Ferrera for
$f=0.75$ is not seen here. 

 In Sec. II we review the formalism for
bounce solutions at finite temperatures. In Sec. III
we discuss our numerical algorithm. In Sec. IV we present our results.
Finally, conclusions are presented in Sec. V.
 
\end {section}
\begin {section}*{II. FINITE TEMPERATURE BOUNCE \\ SOLUTIONS IN FIELD THEORY}

\indent Let us consider a scalar field theory with a Lagrangian density
\begin{equation}
{\mathcal L}(\varphi) = {1 \over 2}({\partial_{\mu}\varphi})^2 - U(\varphi) ,
\end{equation}
 where the potential $U(\varphi)$ has two minima at 
$\varphi_-$ (false vacuum) and $\varphi_+$ (true vacuum). 

 In the semi-classical approximation the barrier tunneling
leads to the appearance of bubbles of a new phase with 
$\varphi = \varphi_+$. To calculate the 
probability 
of such a process in quantum field theory at zero temperature, one should 
first solve the Euclidean equation of motion : 
\begin{equation}
 \partial_{\mu}\partial_{\mu}\varphi = {dU(\varphi) \over d\varphi}
\label{eq:motion},
\end{equation}
with the boundary condition 
$\varphi \to \varphi_-$ as $ \vec x^2+\tau^2 \to \infty$ , 
where $\tau$ is the imaginary time. The probability of tunneling per unit 
time per unit volume will be given by 
\begin{equation}
\Gamma = A \quad e{^{-S_E[\varphi]}} \label{eq:S4},
\end{equation}
where $ S_E[\varphi]$ is the Euclidean action corresponding to the solution 
of Eq.~(\ref{eq:motion})  and given by the following expression :
\begin{equation}
S_E[\varphi] = \int d^4{x} \left[ {1 \over 2} ({\partial\varphi \over 
\partial\tau})^2 + {1 \over 2} ( \nabla\varphi)^2 + U(\varphi) \right] . 
\end{equation}
\indent It is sufficient in most cases to restrict ourselves 
to the O(4) symmetric 
solution $ \varphi(\vec x^2+\tau^2)$, since it is this solution that 
provides the minimum $S_4$ of the action $ S_E[\varphi]$ \cite{Glasser}.
 In this case Eq.~(\ref{eq:motion}) takes the simpler form 
\begin{equation}
{d^2\varphi \over d\rho^2} + {3 \over \rho} {d\varphi \over d\rho} 
= {dU(\varphi) \over d\varphi } , 
\end{equation}
 where $\rho=\sqrt{\vec x^2+\tau^2}$,  
with boundary conditions
\begin{equation} 
\varphi \to \varphi_-  \quad as \quad \rho \to \infty ,\quad 
{d\varphi \over d\rho }= 0 \quad at \quad \rho = 0 .
\end{equation}
\indent Now, let us consider the finite temperature 
case. Following \cite{Linde}, in order 
to extend the above-mentioned results to nonzero temperature, 
$ T \not = 0 $, it is sufficient to remember that quantum statistics of 
bosons (fermions) at $ T \not = 0$ is equivalent to quantum field theory 
in the Euclidean space-time, periodic (anti-periodic) in the ``time'' 
direction with period $ \beta = T^{-1}$. One should use the $T$-dependent 
effective potential $U(\varphi,T)$ instead of the zero-temperature one 
$U(\varphi)=U(\varphi,0)$. Instead of looking for O(4)-symmetric solution 
of Eq.~(\ref{eq:motion}), one should look for O(3)-symmetric 
(with respect to spatial coordinates) solutions, periodic in 
the ``time'' direction with period 
$\beta = T^{-1}$. At sufficiently large temperature compared to the inverse 
of the bubble radius R at $T=0$, the solution is a cylinder whose spatial 
cross section is the O(3)-symmetric bubble of new radius $R(T)$. In this 
case, in the calculation of the action $S_E[\varphi]$, the integration 
over $\tau$ is reduced simply to multiplication by $T^{-1}$, i.e., 
$S_E[\varphi]= T^{-1} \> S_3[\varphi]$, where $S_3[\varphi]$ is 
a three-dimensional action corresponding to the O(3)-symmetric bubble:
\begin{equation}
 S_3[\varphi] = \int d^3 r \left[ {1 \over 2} {(\nabla \varphi)^2 } 
+ U(\varphi,T) \right]  .
\end{equation}
\indent To calculate $S_3(\varphi)$ it is necessary to solve the equation
\begin{equation}
{d^2\varphi \over dr^2} + {2 \over r }{d\varphi \over dr}= 
{dU(\varphi,T) \over d\varphi } \label{eq:S3}
\end{equation}
with boundary conditions 
\begin{equation}
\varphi \to \varphi_- \quad as \quad r \to \infty , \quad
{d\varphi \over dr} = 0 \quad at \quad r=0 .
\end{equation}
where now $r = \sqrt{\vec x^2}$. The complete expression for the 
probability of tunneling per unit time per unit volume in the 
high-temperature limit ($T >> R^{-1}$) is obtained in analogy to 
the one used in \cite{Coleman} and is given by:
\begin{equation}
\Gamma(T) = A(T) \> e^{-S_3[\varphi,T]/T} .
\end{equation}
\indent At intermediate temperature the action will be given, under
the assumption of $O(3)$ symmetry, by
\begin{equation}
S_E[\varphi,T] = 4\pi \int d\tau \int dr \; r^2 \left[ {1 \over 2}{ 
({\partial\varphi \over \partial\tau})^2 + ( \nabla\varphi)^2} 
 + U(\varphi,T) \right] . 
\end{equation}
and the equation of motion will be  
\begin{equation}
{\partial^2\varphi \over \partial\tau^2} +
{\partial^2\varphi \over \partial r^2} + 
{2 \over r }{\partial\varphi \over \partial r} = 
{\partial U(\varphi,T) \over \partial\varphi} , 
\end{equation}
with boundary conditions 
\begin{equation}
\varphi \to \varphi_- \quad as \quad r \to \infty ,\quad 
\partial\varphi/\partial\tau = 0 \quad at \quad \tau = \pm \beta/2,0
\end{equation}
where $\beta$ is the period of the solution.

\end {section}
\begin {section}*{III. ALGORITHM}

The aim of this paper is to investigate the nature of the
phase transition for $\varphi^4$ theory with a $\varphi^3$ symmetry-breaking 
term and compare
our results with those of Ferrera \cite{Ferrera} who discussed a $\varphi$
symmetry-breaking term.

 We start with the following Euclidean action inspired by recent work
on temperature-dependent corrections to the tree level potential
\cite{Dine} :
\begin{equation}
S_E = 4\,\pi \int d\tau \int dr \; r^2 \left[ {1 \over 2}{ 
({\partial\varphi \over \partial\tau})^2 + ( \nabla\varphi)^2} 
 + U(\varphi) \right] . 
\end{equation}
where $U(\varphi)$ is 
\begin{equation}
 U(\varphi)= {\lambda \over 2}( \varphi^2 - \mu^2)^2 -F \varphi^3 \label{eq:pot}.  
\end{equation}  .
We rescale the Lagrangian density as
\begin{equation}
\varphi \to \varphi / \mu ,\quad r \to r\sqrt{\lambda \mu^2} ,\quad f=
{F \over {\lambda \mu}} 
\end{equation}
to get
\begin{equation}
S_E = {4\,\pi \over \lambda} \int d\tau \int dr \; r^2 \left[ {1 \over 2}{ 
({\partial\varphi \over \partial\tau})^2 + ( \nabla\varphi)^2} 
 + {1 \over 2}( \varphi^2 - 1)^2 -f \varphi^3  \right] . 
\end{equation}
 The only free parameter in the Lagrangian is $f$. Taking different
values of $f$ will give us different shapes of the potential. Hence
we will be able to explore the nature of the phase transition for
thin as well as for thick walls. The equation of motion now
becomes
\begin{equation}
{\partial^2\varphi \over \partial\tau^2} +{\partial^2\varphi \over \partial
r^2} + {2 \over r }{\partial\varphi \over \partial r} =
2(\varphi^3-\varphi)-3f\varphi^2 \label{eq1:motion}, 
\end{equation}
with boundary conditions 
\begin{equation}
\varphi \to \varphi_- \quad as \quad r \to \infty ,\quad 
\partial\varphi/\partial\tau = 0 \quad at \quad \tau = \pm \beta/2,0 .
\end{equation}
\par
To solve Eq.~(\ref{eq1:motion}) we have used a multigrid
algorithm. Multigrid methods were introduced in the 1970s by Brandt   
\cite{Brandt}. More details are available elsewhere 
\cite{Press,Hackbusch}.

 We have used full weighting injection for the
projection operator and second-order polynomial interpolation. For the
thin wall we have used fourth order polynomial interpolation since
more  points are coupled to each other than in the thick-wall case. Since
we used the multigrid algorithm (not the full multigrid \cite{Press}), we have
to start from the finest grid and go down to coarse grids and then up
and down till the solution converges. Our finest grid has $513 \times
513$ points. We have restricted  ourselves to three grids ($513 \times
513, \; 257 \times 257 , \;133 \times 133 $).

 As we can see Eq.~(\ref{eq1:motion}) has a singularity 
at $r=0$. One way to handle it by using  an arbitrary regularization
$1/r \to {1/(r + \epsilon)}$ with $\epsilon$ chosen
suitably small. This is the method used by Ferrera \cite{private}. 
We use an alternate method based on L'Hospital's rule, replacing 
${(1/r)\partial\varphi/\partial r}$ at $r=0$ by
${\partial^2\varphi/\partial r^2}$.
 
 We have started with the zero temperature solution as an initial guess
solution, as the zero-temperature equation of motion is 
one-dimensional and straightforward to solve \cite{Hatem}. We used
this one-dimensional solution after putting
it in two-dimensional form. After computing the zero-temperature
solution by the multigrid method we used it as a guess solution for our first
finite-temperature solution. To compute the next finite-temperature solution we
used the previous one as input, and so on. We followed this procedure
(i.e., continuation in temperature) to cover the whole range of 
temperatures.

At zero temperature we have used point Jacobi relaxation
and at finite temperature we have used line Newton Jacobi relaxation 
in the $\tau$ direction because as the temperature is increased we have to
reduce the step size in the $\tau$ direction and this means that
more points in the $\tau$ direction will be coupled than in the other
direction. 

 To solve the equation of motion, we need the value $\varphi_{Tc}$ of
the bounce at the center. For this, we chose a trial value $\varphi_c$ and
kept it fixed through the relaxation process (since otherwise the process is
bound to diverge \cite{Ferrera}). In the zero- and high-temperature
limits, $\varphi{T_c}$ was fixed very accurately by solving the
one-dimensional equation of motion using a predictor-corrector
method and shooting in $\varphi_c$ \cite{Hatem}. For 
multigrid, we have found the same behavior as Ferrera
\cite{Ferrera}, i.e. if we give too high value of $\varphi_c$ the multigrid
solution diverges while for low enough values it converges. If 
$\varphi_{c} - \varphi_{Tc} \sim \Delta$, the 
solution converges to an overall 
accuracy level of order $\Delta$. We found  $\Delta \sim 0.1\%$
which corresponds to a low level of error for
the solution. The shooting in $\varphi_{c}$, together with the
variable amount of relaxation required, makes the program very
CPU-intensive. A typical running time for a single value of temperature
is about an hour and a half on an 80586 Pentium machine running at 133 MHz.

 A drawback of the multigrid method is that we cannot find the solution
at an arbitrary value of temperature. We have a solution only at discrete
values of $\beta$. A constant step size in $\beta$ implies the interval
between successive temperatures increases as we go to higher temperatures. On
the other hand, the $\beta$ step cannot be made very small because of the
reasons stated above.

Finally, to compute the action for each solution we have used 
a two-dimensional Simpson method that can be found in \cite{Abramowitz}.

\end {section}
\begin{section}*{IV. RESULTS}

We have used the algorithm described in the above section 
for solving the equation of motion 
\begin{equation}
{\partial^2\varphi \over \partial\tau^2} +{\partial^2\varphi \over \partial
r^2} + {2 \over r }{\partial\varphi \over \partial r} =
2(\varphi^3-\varphi)-3f\varphi^2 .
\end{equation}
To be able to compare our results with Ferrera \cite{Ferrera}, we have chosen
the same three values of the parameter $f$. The first value is $f = 0.25$.
From \cite{Hatem}, $f = 0.25$ is within the domain of validity of the TWA 
Fig. 1 represents the shape of the potential for $f = 0.25$.
If we compare the action of the zero- and high-temperature solutions with
the one-dimensional one, we find a difference of order
$0.2\%$. At intermediate temperatures the highest difference is about
$0.9\%$. Fig. 2 shows a plot of $S(T) ~vs~ \beta$. From it,
the value of $\beta$ at which the transition takes place is 
approximately 21.15 in our dimensionless units.
Garriga \cite{Garriga} has shown that, in the TWA, the period 
at which the transition takes place is given by
\begin{equation}
\beta_c = {{27\pi} \over 32 }R \label{eq:T_c} ,
\end{equation}
where $R$ is the radius of the one-dimensional zero-temperature solution.
We have measured the radius from the center of the bounce to the
middle of the wall where the absolute first derivative of the field is
maximum. Using Eq.~(\ref{eq:T_c}), we obtain
a value of $\beta_c \approx 21.205$. The difference between the 
numerical result and the
analytic one  is about $0.2\%$, which lies within
our estimation of errors. Clearly it is a first-order phase
transition and this will always happen as long as the thickness of the
wall is much smaller than the radius of the nucleated bubble.
Fig. 3(a) shows the zero-temperature solution. We can see that the
wall thickness is much smaller than the radius. Fig. 3(b) shows 
the solution at
$\beta =25.3$. We can see that the two bubbles are far from touching
each other. Fig. 3(c) shows the solution at $\beta=\beta_c$. In this 
case we can see that the
two bubbles just fit within the period. Finally, Fig. 3(d) shows
the solution at $\beta\leq\beta_c$.

The second value is at $f=0.55$. Fig. 4 shows the shape of the
potential. It is clear from the figure that the TWA is not valid since the
energy difference between the false and true vacua is larger than 
the hump at the origin. Fig. 5 shows a plot of $S(T) ~vs~ \beta$. From 
the figure, we can see that the
action is constant till $\beta=14.4$ and after that it starts decreasing
slowly till it matches with the $\beta S_3$ curve, where $S_3$ is the
high temperature action. The transition point is at $\beta=9.3$. If $f
=0.55$ were within the TWA, then the transition point would
be at $\beta{_\star}=9.9$. We can see that the transition point is at a period
less than $\beta{_\star}$.
Fig. 6(a) shows the zero-temperature
solution. Fig. 6(b) shows the solution at $\beta=14.4$, where 
the two bubbles have not started touching each other. Hence we
have constant action till that period.
Fig. 6(c) shows the solution at $\beta=9.3$. We can see that the
solution looks like a wiggly cylinder but it is far from
a static one, so we still do have a first-order phase transition, 
but more weakly than for $f = 0.25$. Finally, Fig. 6(e) shows the 
solution at $\beta \leq 9.3$.

 Finally, Fig. 7 shows the shape of the potential for $f =
0.75$. Fig. 8 shows a plot of $S(T) ~vs~ \beta$. From the figure we 
can see that the
action is constant till $\beta=10.3$. After that it starts decreasing
till it crosses the $\beta S_3$ curve at $\beta=6.25$. Again, if $f =0.75$
were within the TWA, then the crossing point would be at
$\beta{_\star}=7.45$ . The transition point is at $\beta=6.25$, 
which is again less than $\beta{_\star}$. Fig. 9(a) shows the
zero-temperature solution. Fig. 9(b) shows the solution at $\beta=10.3$;
we can see that wiggly cylinder solution has not started forming
yet. Fig 9(c) shows the solution at the transition point
$\beta=6.25$. We can see that the wiggly cylinders are formed and still 
far from a static solution. Hence we have a first-order phase
transition but more weakly than for $f=0.55$. Finally, Fig. 9(d) shows the
solution at $\beta \leq 6.25$. 

\end {section}
\begin{section}*{V. CONCLUSIONS}

In this paper we have tried to find out whether the phase 
transition from the
quantum regime to the high-temperature regime is of first  or
second order for $\varphi^4$ theory with a $\varphi^3$ symmetry-breaking
term. We have investigated three values of $f$. We find that for  
$f = 0.25$ we have a first order phase transition. For $f= 0.55$, the
transition is still first-order but weaker than for
$f=0.25$. Another feature is that the transition point $\beta_c$ is 
less than the point $\beta_\star$
where the zero-temperature solution will match the high temperature
one. The final value is $f=0.75$. Here we have a weak first order
phase transition. Again $\beta_c  < \beta_\star$. In this case the 
departure is more than for $f =0.55$.
So, up to $f=0.75$ we have a first-order phase transition, which is
different from what Ferrera \cite{Ferrera} found for $f= 0.75$. We conclude
that, for the potential with a $\varphi^3$ symmetry-breaking term, a true
second-order phase transition is not seen even at
$f=0.75$. This can be understood as follows. A second-order phase 
transition only occurs when the hump in the potential disappears.
In \cite{Hatem} we found that for $\varphi$ symmetry breaking 
the hump in $U(\varphi$) disappears for $f > 0.77$, while for
$\varphi^3$ it disappears only asymptotically in $f$. 
The two forms of the potential
can be mathematically transformed into each other \cite{Adams}. However, this
obscures the physical meaning of the asymmetry coefficient $f$, so
we work with the untransformed potential.

We find that the departure of the transition point ($\beta_c$) 
from the point where
O(4) and O(3) match ($\beta_\star$) is higher for higher values of $f$.
This is expected since, in the domain of validity of the TWA, the surface
tension of the bubble is independent of the temperature \cite{Hatem}; hence we 
must have $\beta_c = \beta_\star$. Beyond the TWA, on the other hand, we can
expect departure from this equality.

 Finally, in this paper we have studied a zero-temperature potential motivated
by the finite-temperature effective potential \cite{Dine}. We have 
incorporated the temperature only in the boundary conditions. In an 
exact calculation, of course, we should use the full temperature-dependent
effective potential. However, we expect that there will be a range of
temperatures and parameter values for which the potential 
of Eq.~(\ref{eq:pot}) will be a good approximation to the full potential.
Hence, our conclusions about the nature of the transition to the thermal
hopping may have relevance to actual physical situations, in particular 
inflationary cosmology and baryogenesis.

\end{section}
\begin{section}*{ACKNOWLEDGEMENTS}

This work is part of a project (No. SP/S2/K--06/91) funded by the
Department of Science and Technology, Government of India.  
 We would like to thank Antonio Ferrera for insightful comments on
the numerical techniques.
 H.W. thanks the University Grants Commission, New Delhi,
for a fellowship.  

\end {section}
\newpage
\bibliography{plain}

\begin{section}*{Figure Caption}

FIG. 1. Shape of the potential for $f=0.25$.

FIG. 2. Temperature dependence of the Euclidean action:
 $S(T) vs \beta$ for $f=0.25$.

FIG. 3. Bounce solution for $f=0.25$. $(a)$ $T=0$, $(b)$ $\beta=25.3$,

$(c)$ $\beta=21.15$, and $(d)$ high temperature.

FIG. 4. Shape of the potential for $f=0.55$.
 
FIG. 5. Temperature dependence of the Euclidean action:
 $S(T) vs \beta$ for $f=0.55$.

FIG. 6. Bounce solution for $f=0.55$. $(a)$ $T=0$, $(b)$ $\beta=14.4$,

$(c)$ $\beta=9.3$, and $(d)$ high temperature.

FIG. 7. Shape of the potential for $f=0.75$.
 
FIG. 8. Temperature dependence of the Euclidean action:
 $S(T) vs \beta$ for $f=0.75$.

FIG. 9. Bounce solution for $f=0.75$. $(a)$ $T=0$, $(b)$ $\beta=10.3$,

$(c)$ $\beta=6.25$, and $(d)$ high temperature. 

\end{section}
\newpage
\begin{figure}[ht]
\vskip 15truecm

\includegraphics{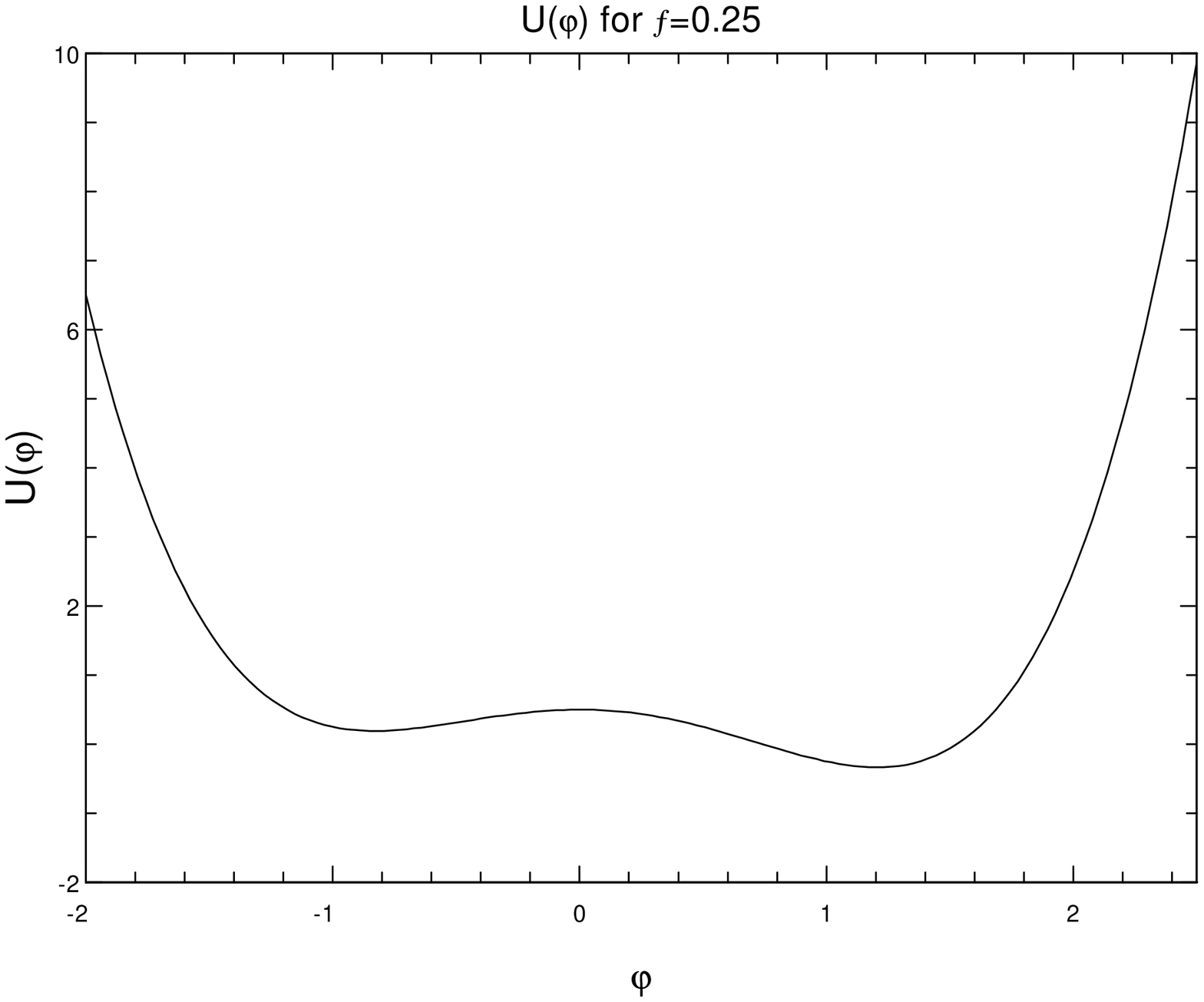} 

\caption{}
\end{figure}
\newpage
\begin{figure}[ht]
\vskip 15truecm

 \includegraphics{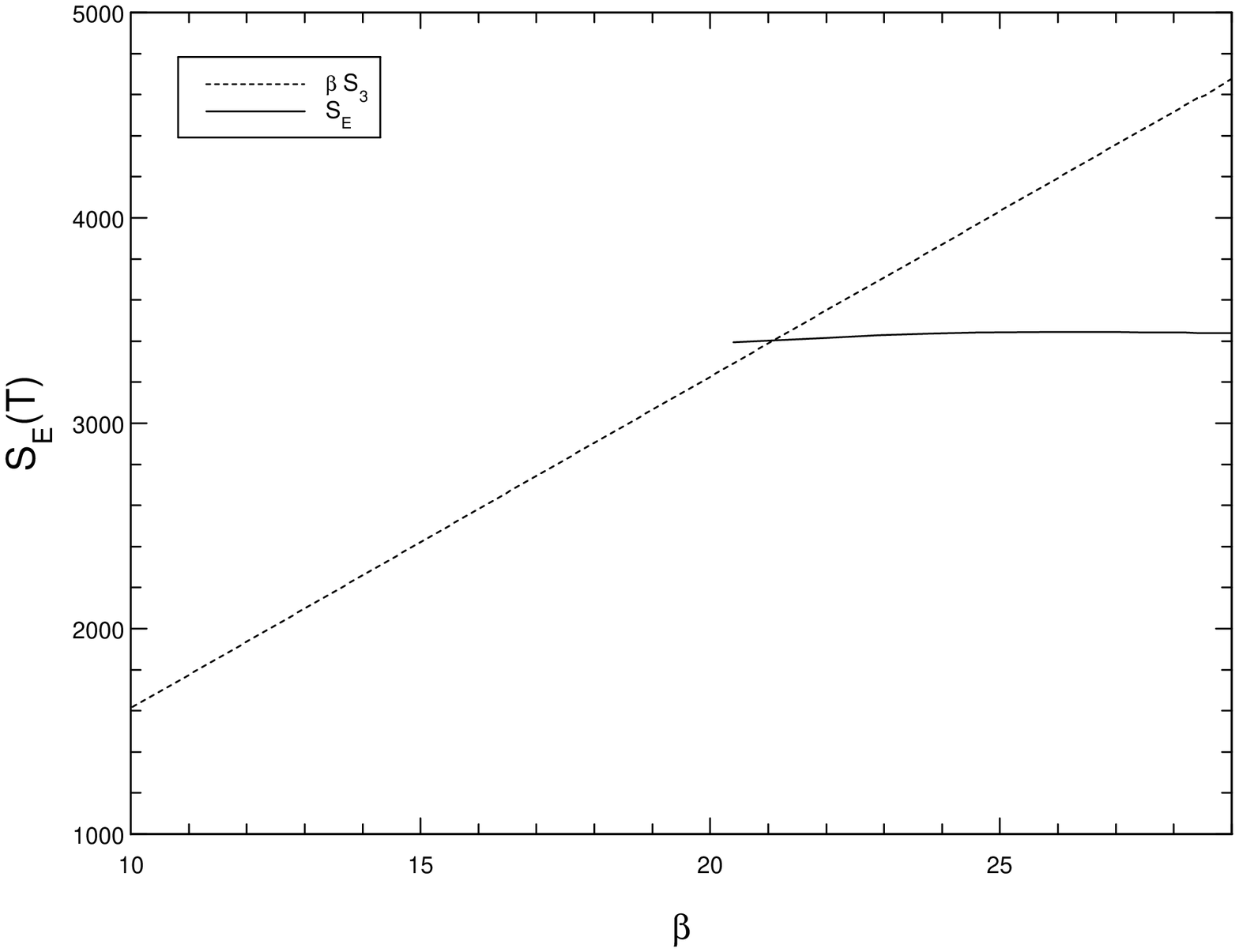} 

\caption{}
\end{figure}
\newpage
\begin{figure}[ht]
\vskip 20truecm

 \includegraphics{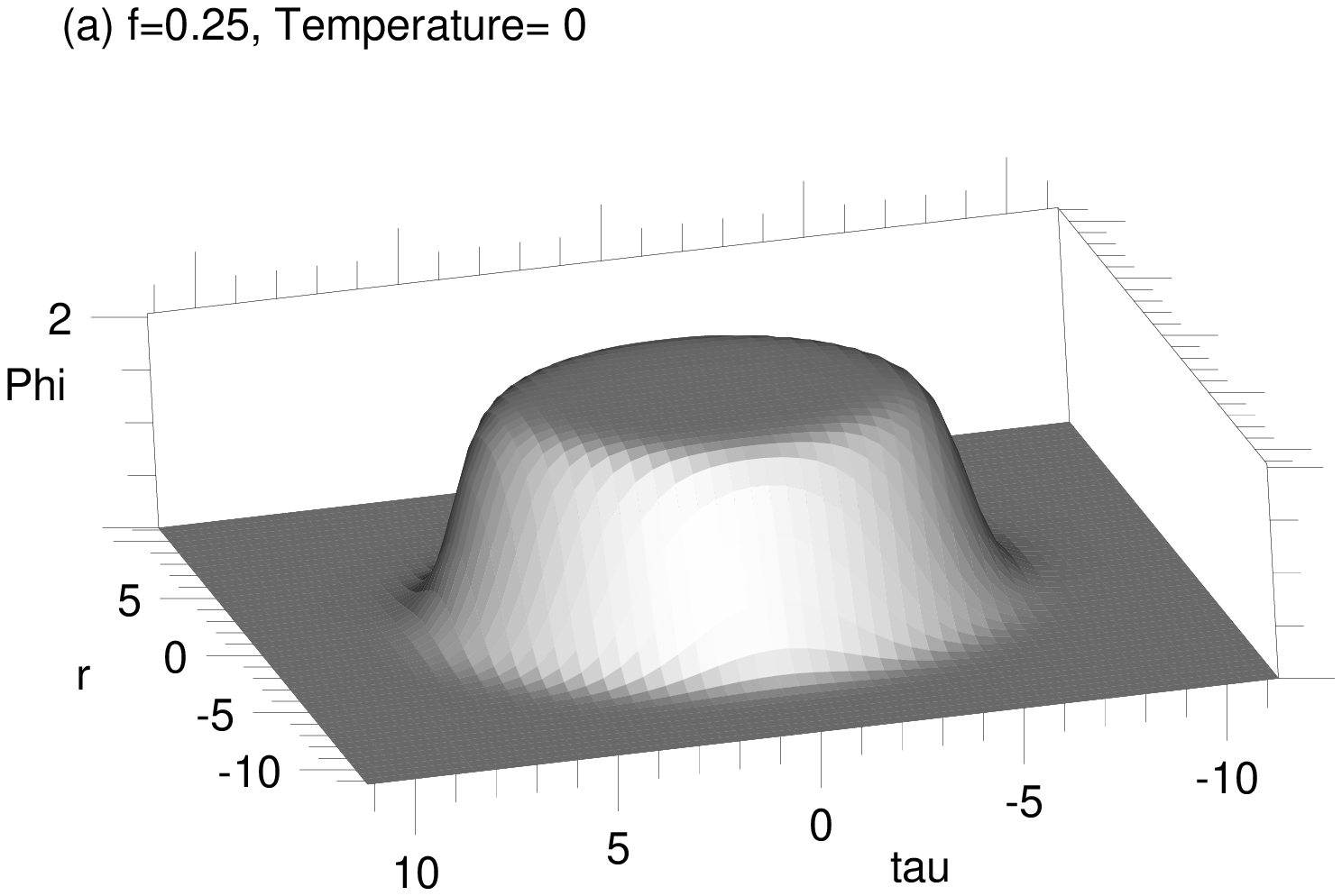} 

\caption{}
\begin{center}
Figure: 3 (a)
\end{center}
\end{figure}
%
\begin{figure}[ht]
\vskip 15truecm

 \includegraphics{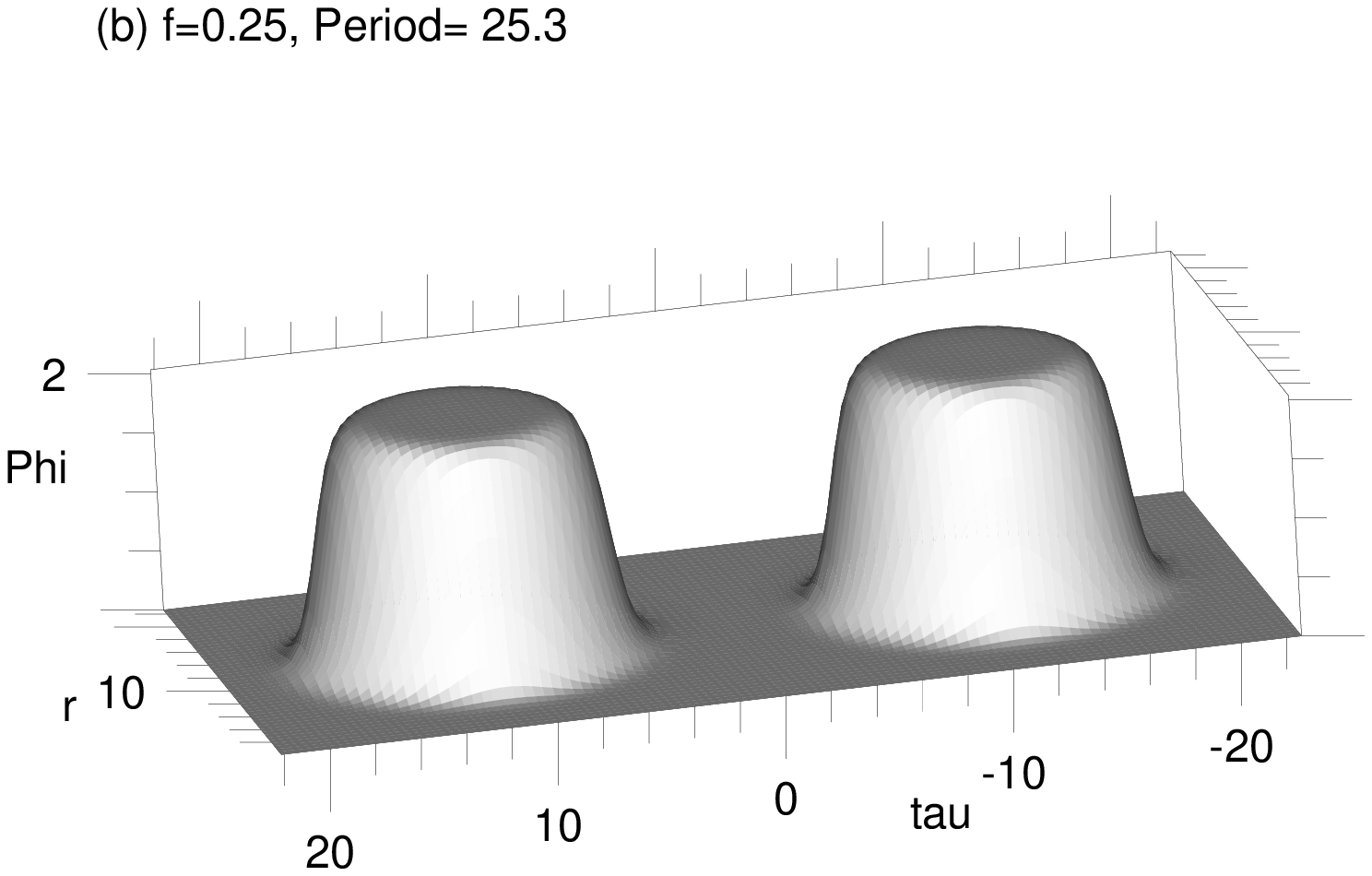} 

\begin{center}
Figure: 3 (b)
\end{center}
\end{figure}
\newpage
\begin{figure}[ht]
\vskip 15truecm

\includegraphics{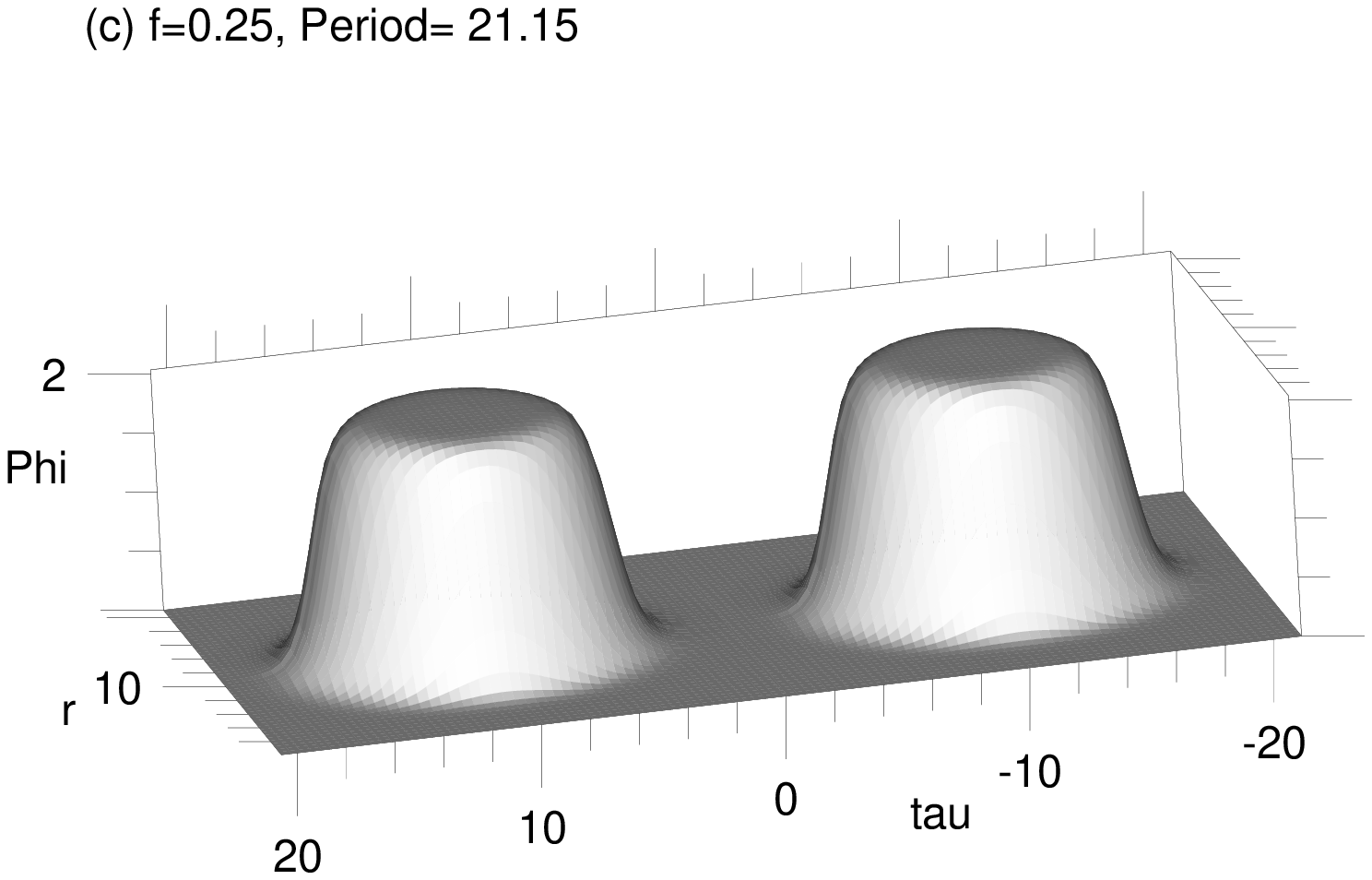} 

\begin{center}
Figure: 3 (c)
\end{center}
\end{figure}
\newpage
\begin{figure}[ht]
\vskip 15truecm

 \includegraphics{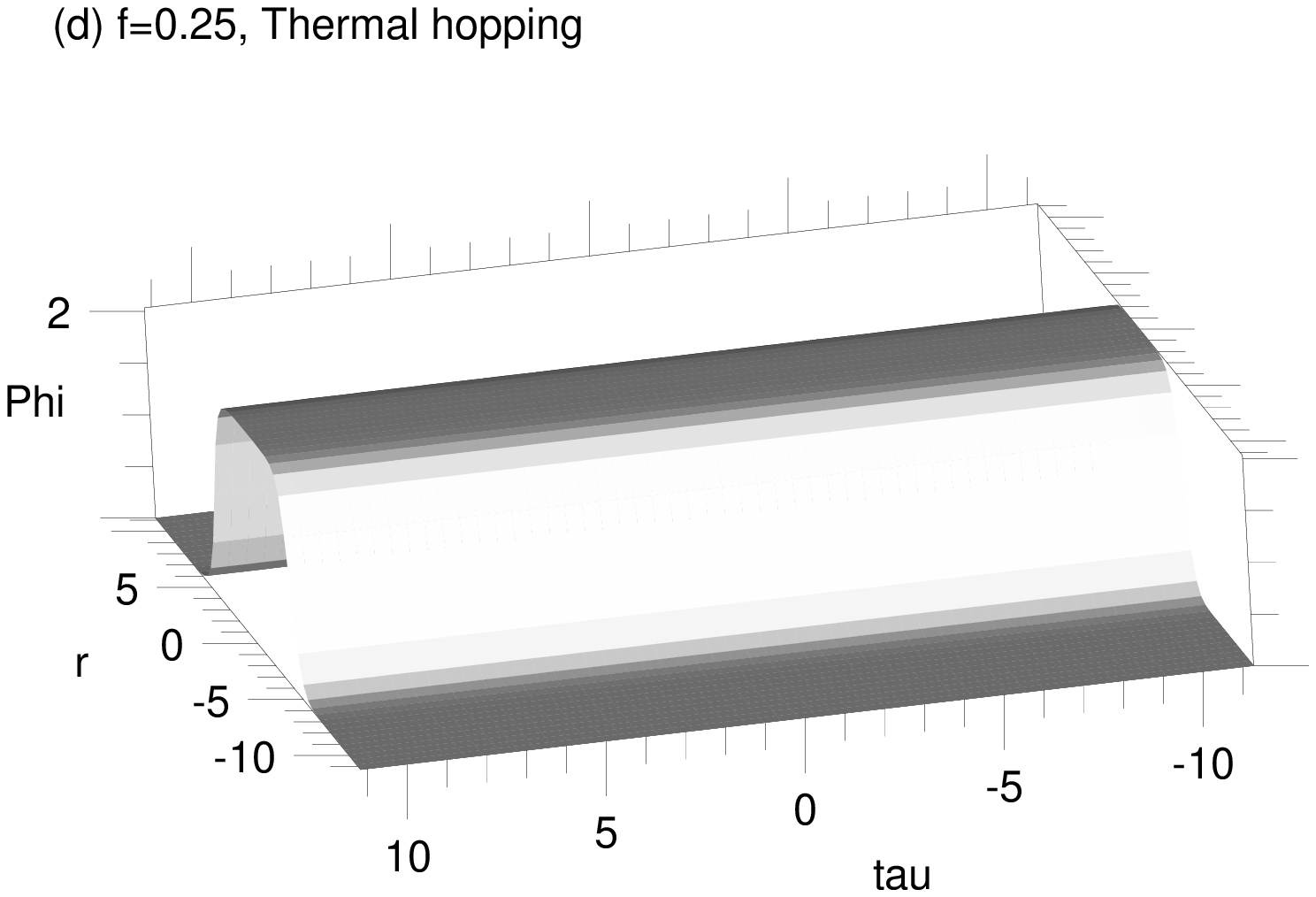} 

\begin{center}
Figure: 3 (d)
\end{center}
\end{figure}
\newpage
\begin{figure}[ht]
\vskip 15truecm

 \includegraphics{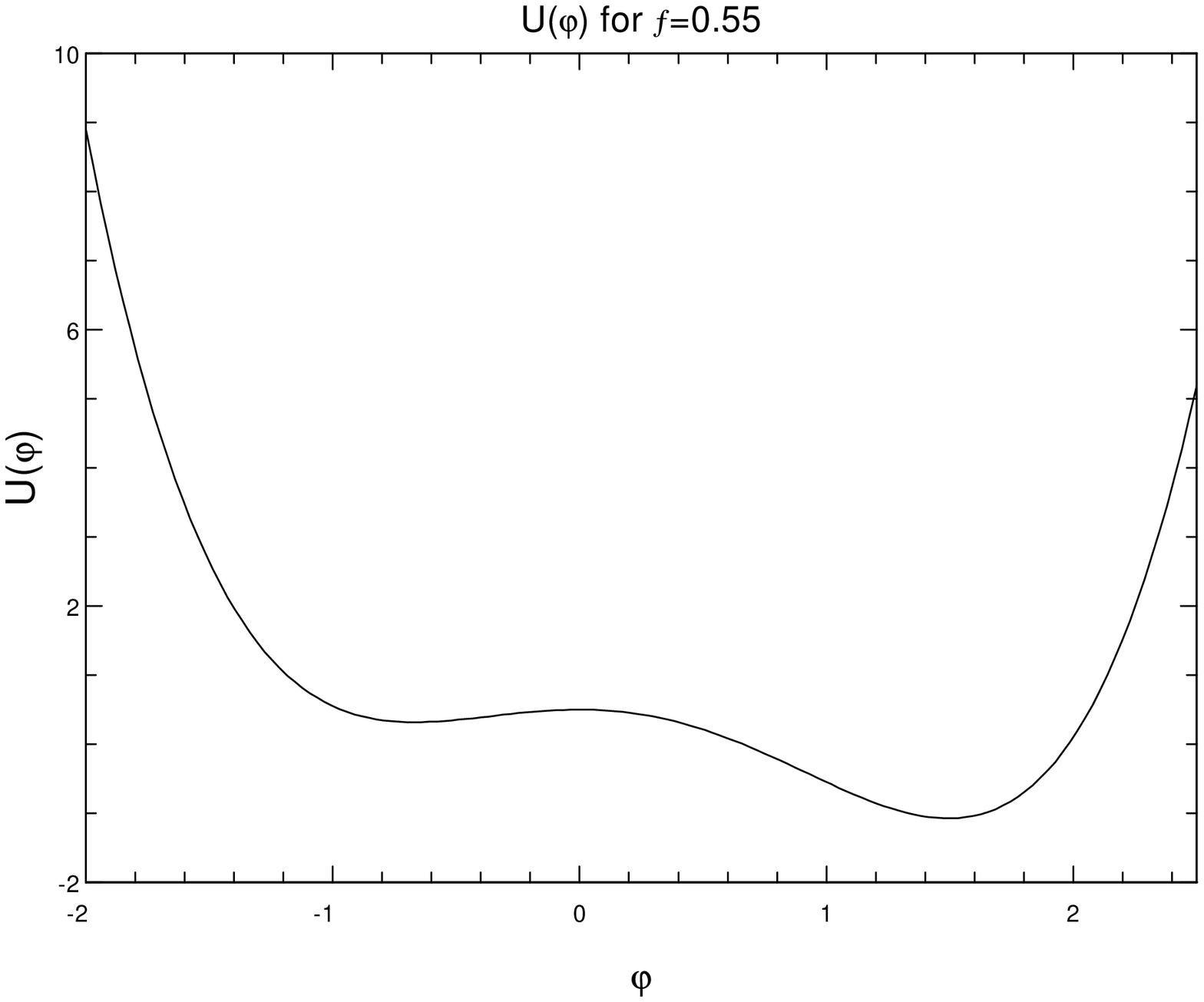} 

\caption{}
\end{figure}
\newpage
\begin{figure}[ht]
\vskip 15truecm

 \includegraphics{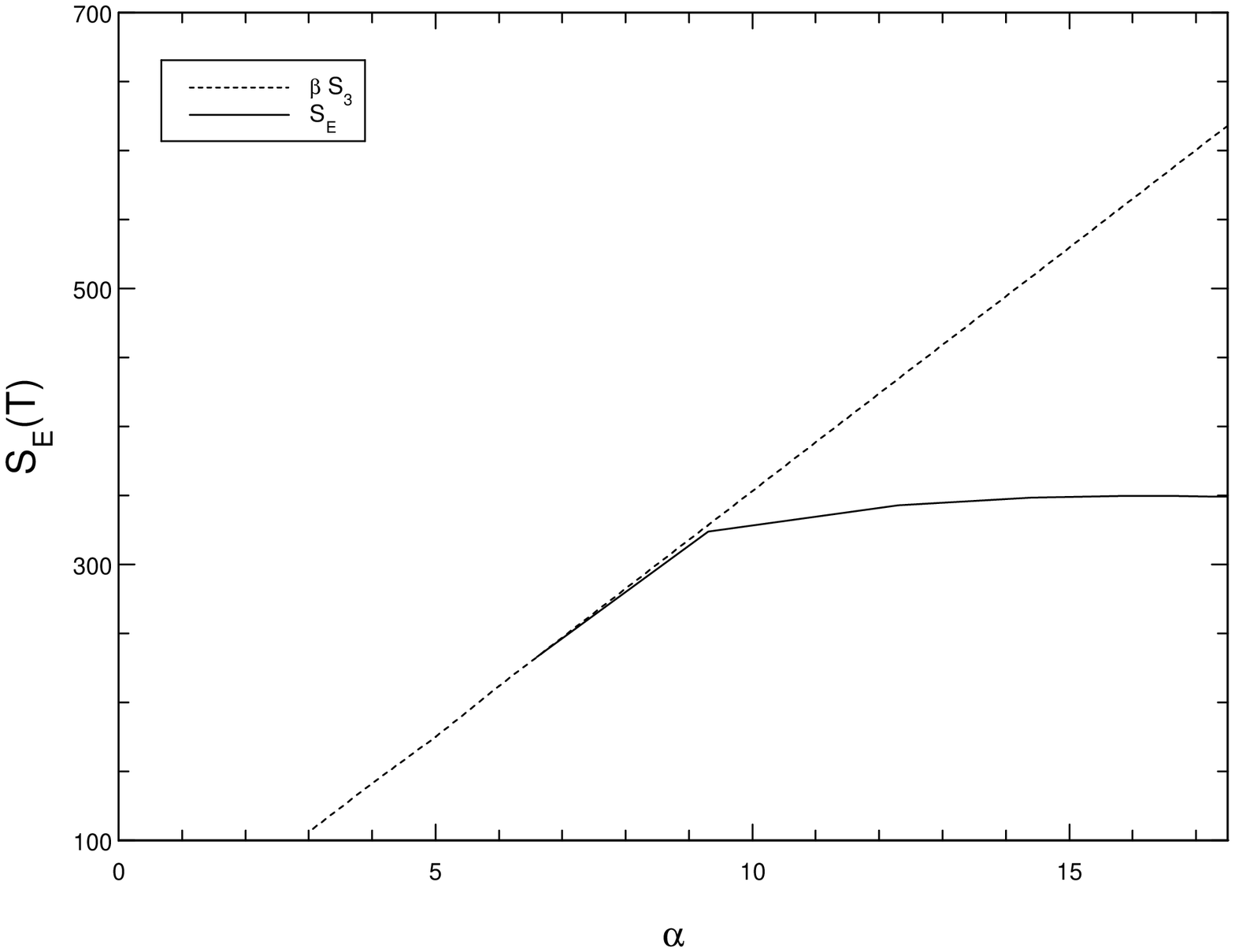} 

\caption{}
\end{figure}
\newpage
\begin{figure}[ht]
\vskip 15truecm

\includegraphics{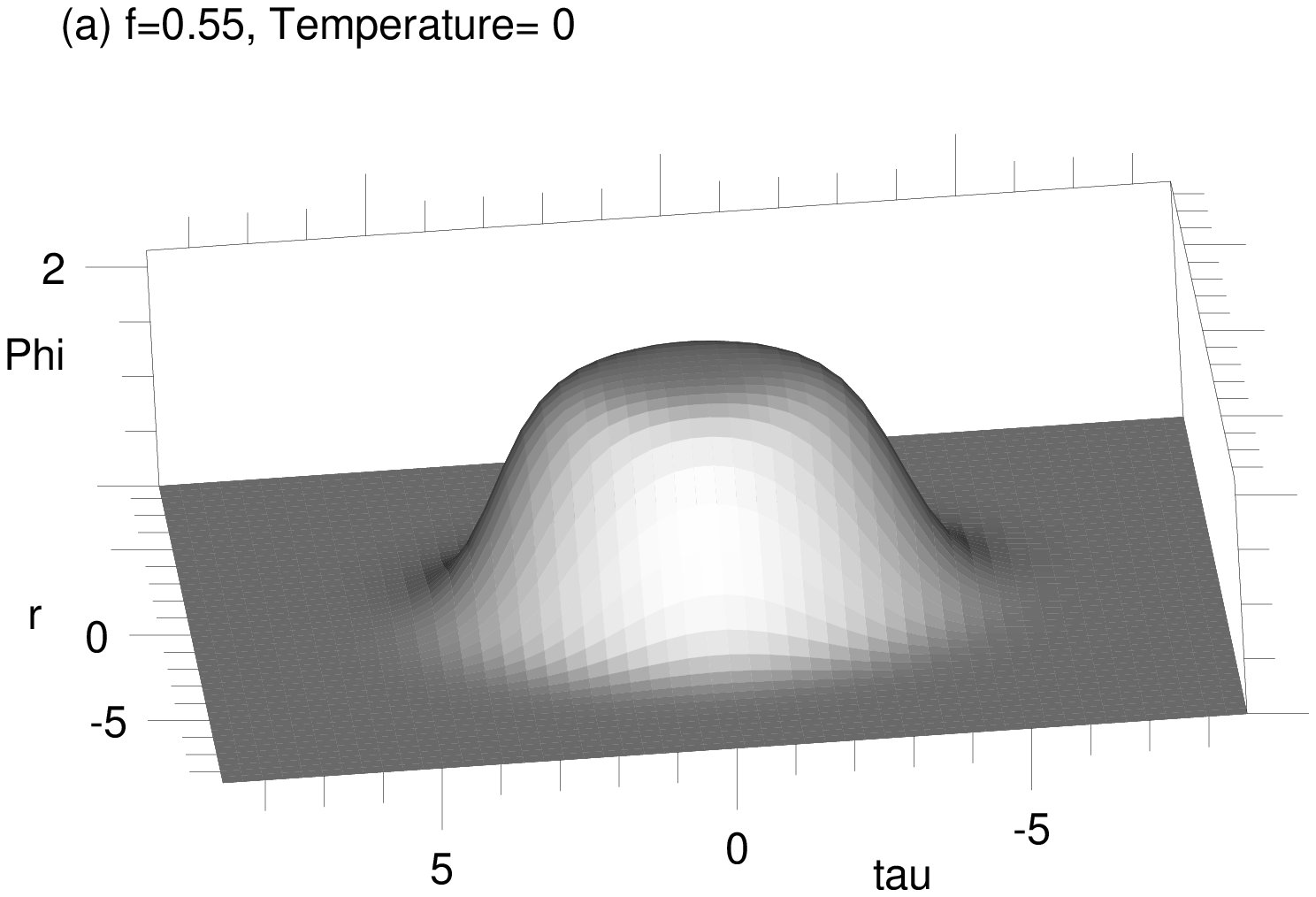} 

\caption{}
\begin{center}
Figure: 6 (a)
\end{center}
\end{figure}
\newpage
\begin{figure}[ht]
\vskip 15truecm

 \includegraphics{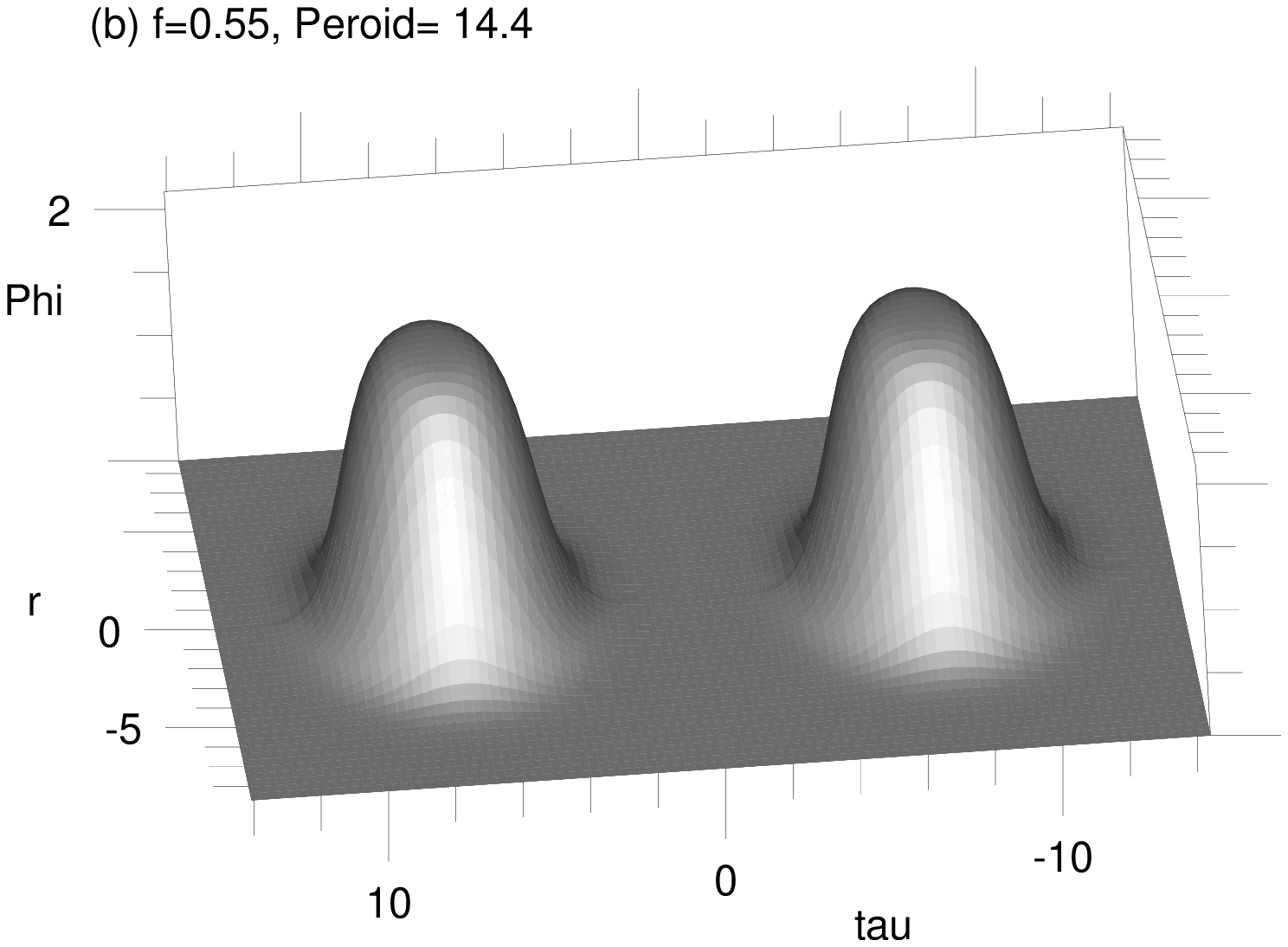} 

\begin{center}
Figure: 6 (b)
\end{center}
\end{figure}
\newpage
\begin{figure}[ht]
\vskip 15truecm

 \includegraphics{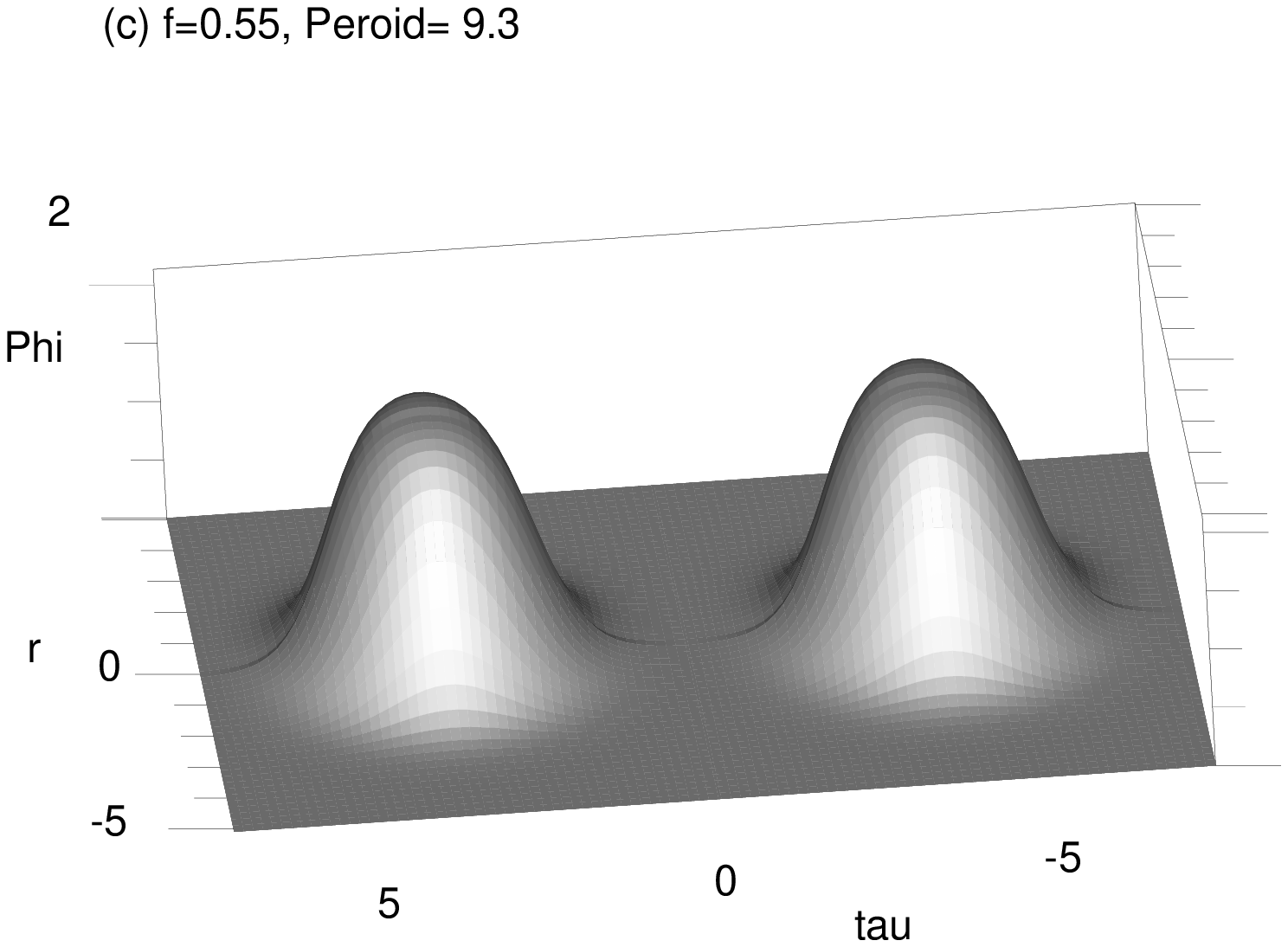} 

\begin{center}
Figure: 6 (c)
\end{center}
\end{figure}
\newpage
\begin{figure}[ht]
\vskip 15truecm

 \includegraphics{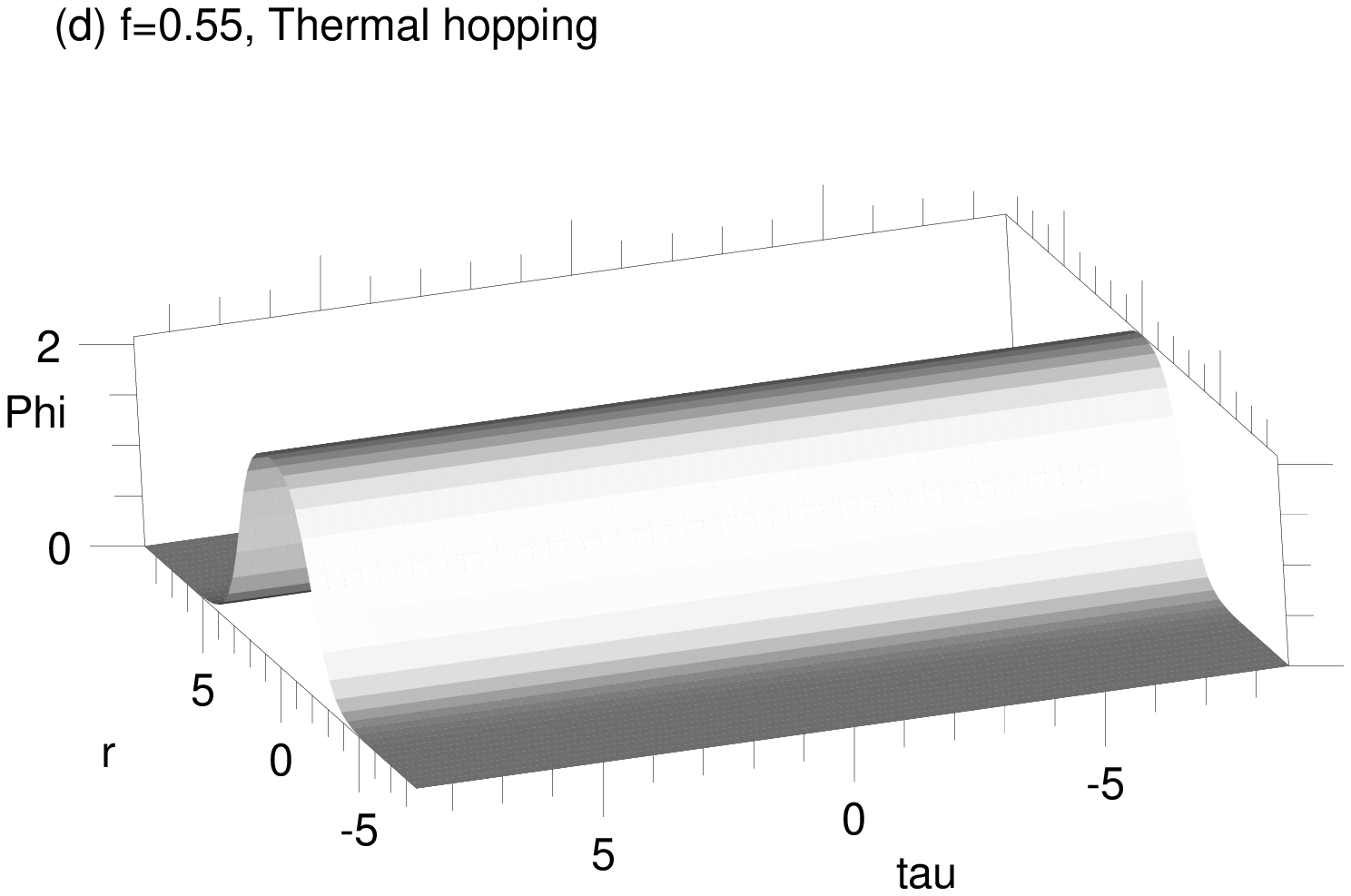} 

\begin{center}
Figure: 6 (d)
\end{center}
\end{figure}
\newpage
\begin{figure}[ht]
\vskip 15truecm

\includegraphics{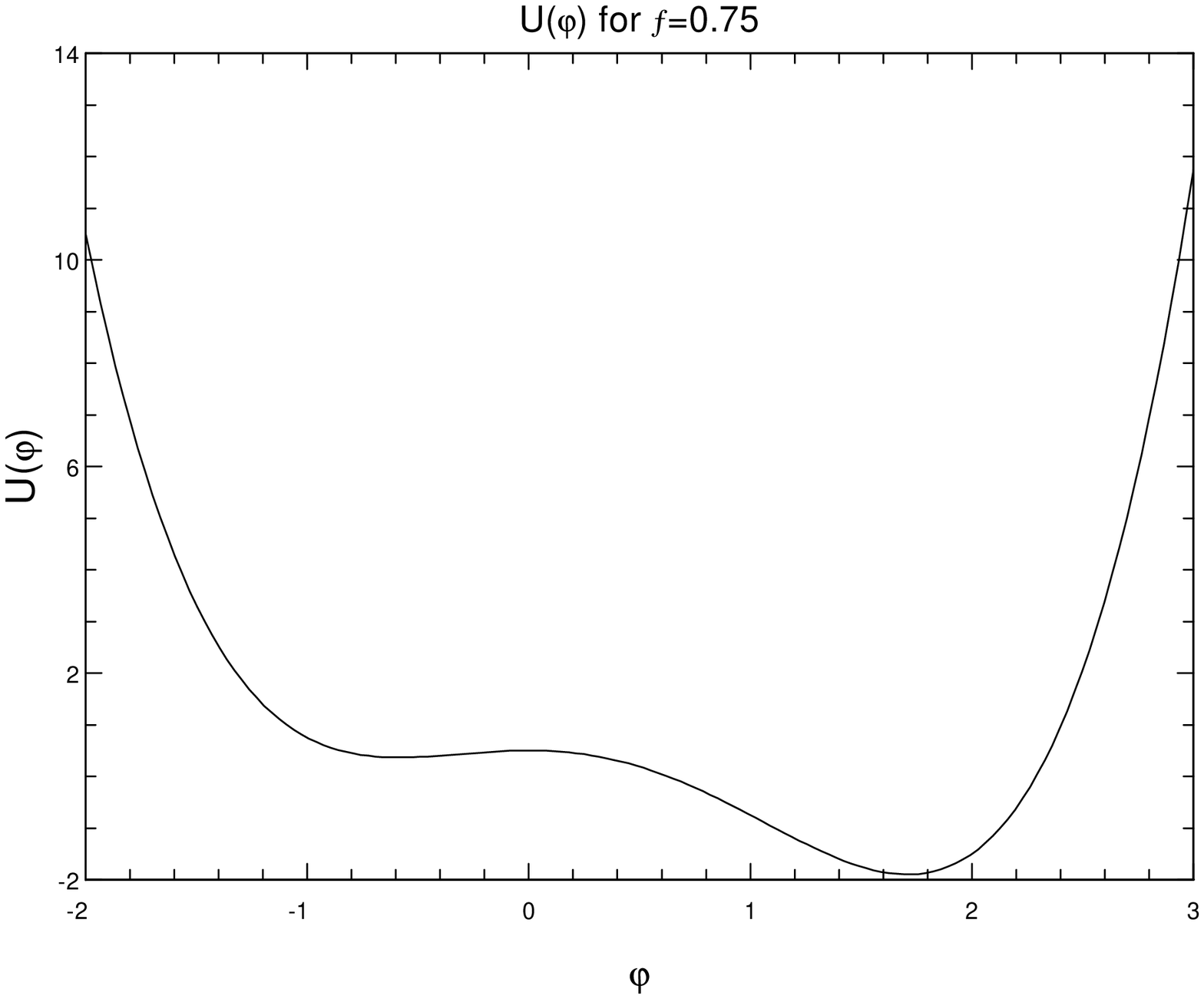} 

\caption{}
\end{figure}
\newpage
\begin{figure}[ht]
\vskip 15truecm

 \includegraphics{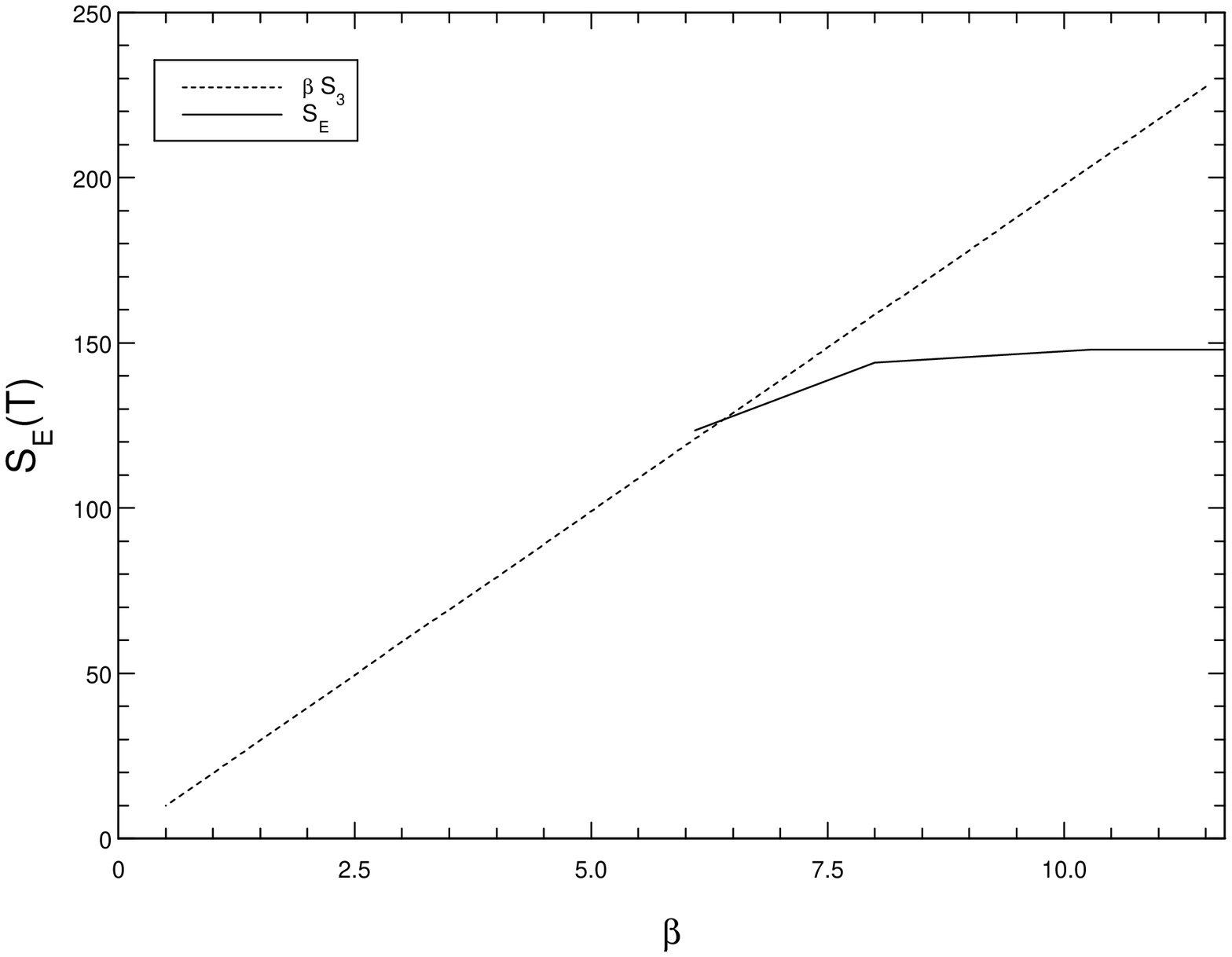} 

\caption{}
\end{figure}
\newpage
\begin{figure}[ht]
\vskip 15truecm

 \includegraphics{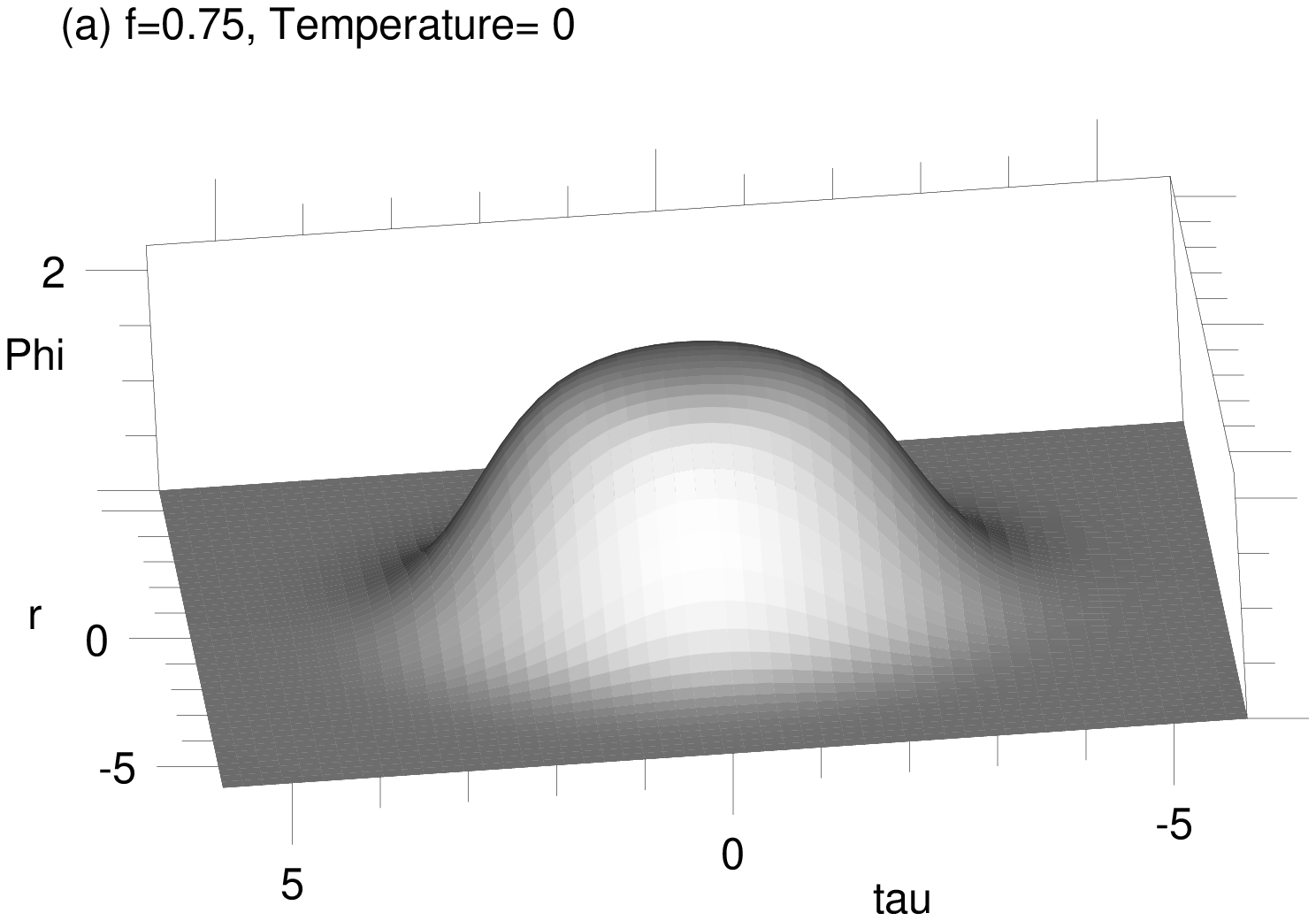} 

\caption{}
\begin{center}
Figure: 9 (a)
\end{center}
\end{figure}
\newpage
\begin{figure}[ht]
\vskip 15truecm

 \includegraphics{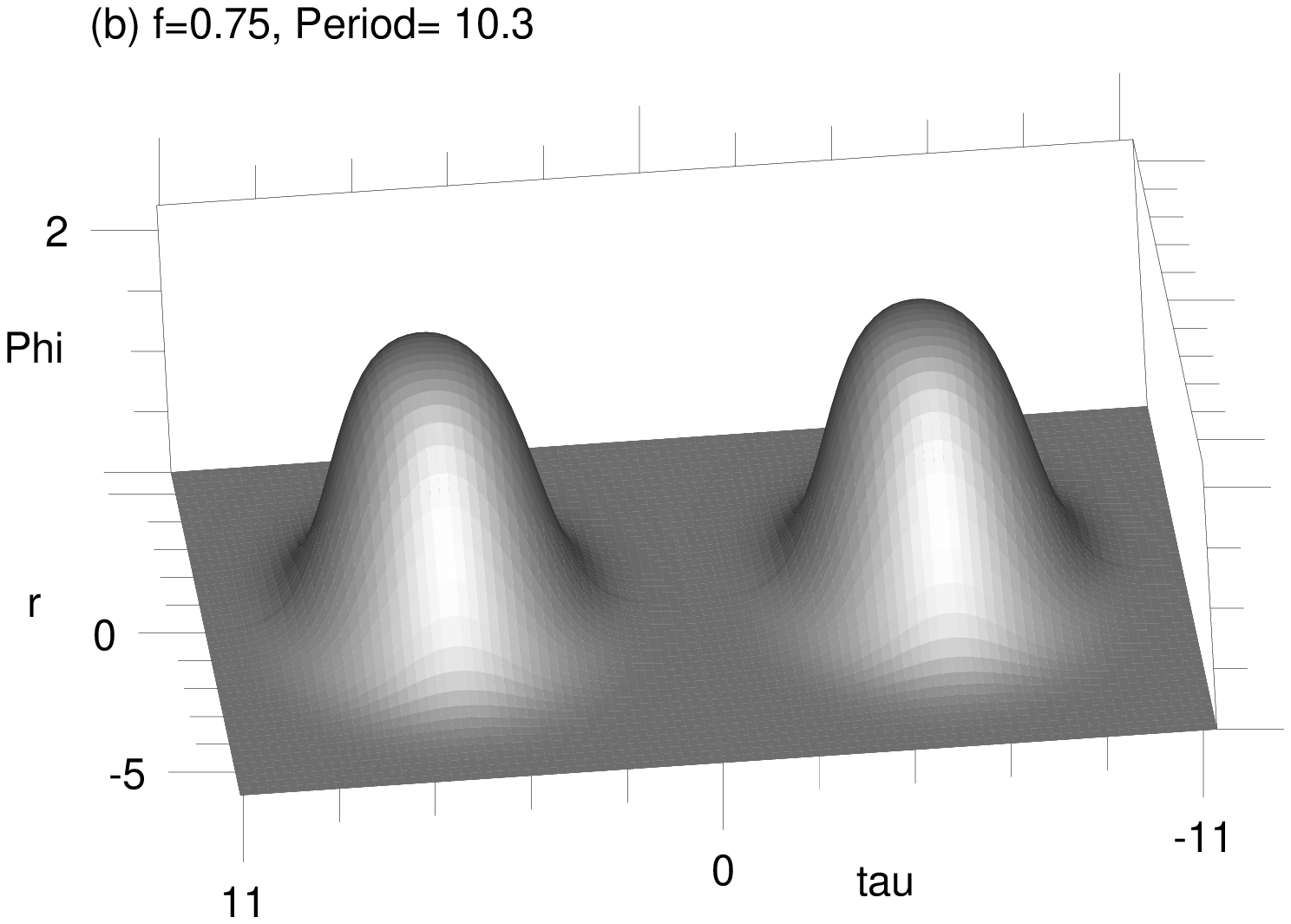} 

\begin{center}
Figure: 9 (b)
\end{center}
\end{figure}
\newpage
\begin{figure}[ht]
\vskip 15truecm

\includegraphics{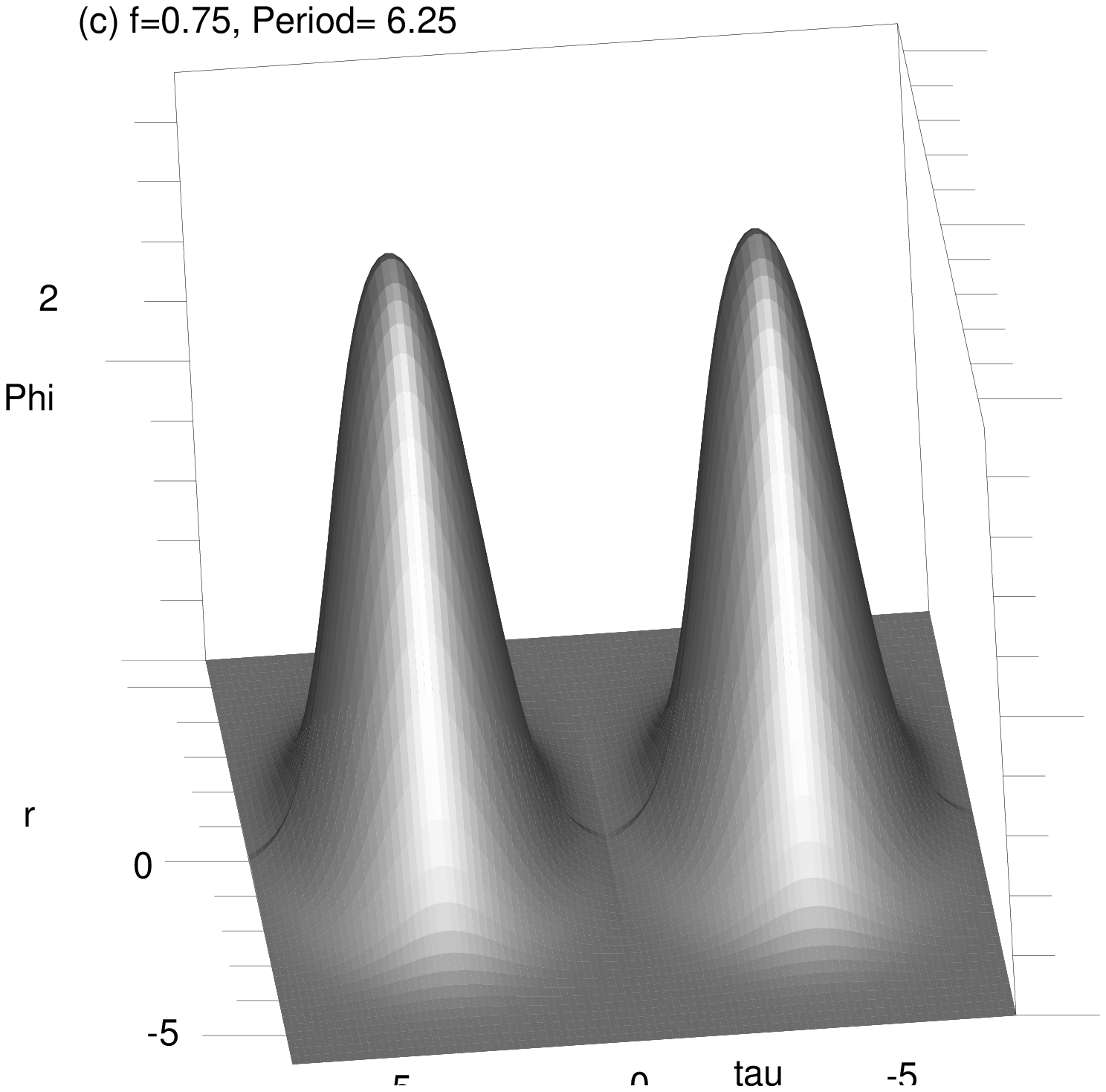} 

\begin{center}
Figure: 9 (c)
\end{center}
\end{figure}
\newpage
\begin{figure}[ht]
\vskip 15truecm

 \includegraphics{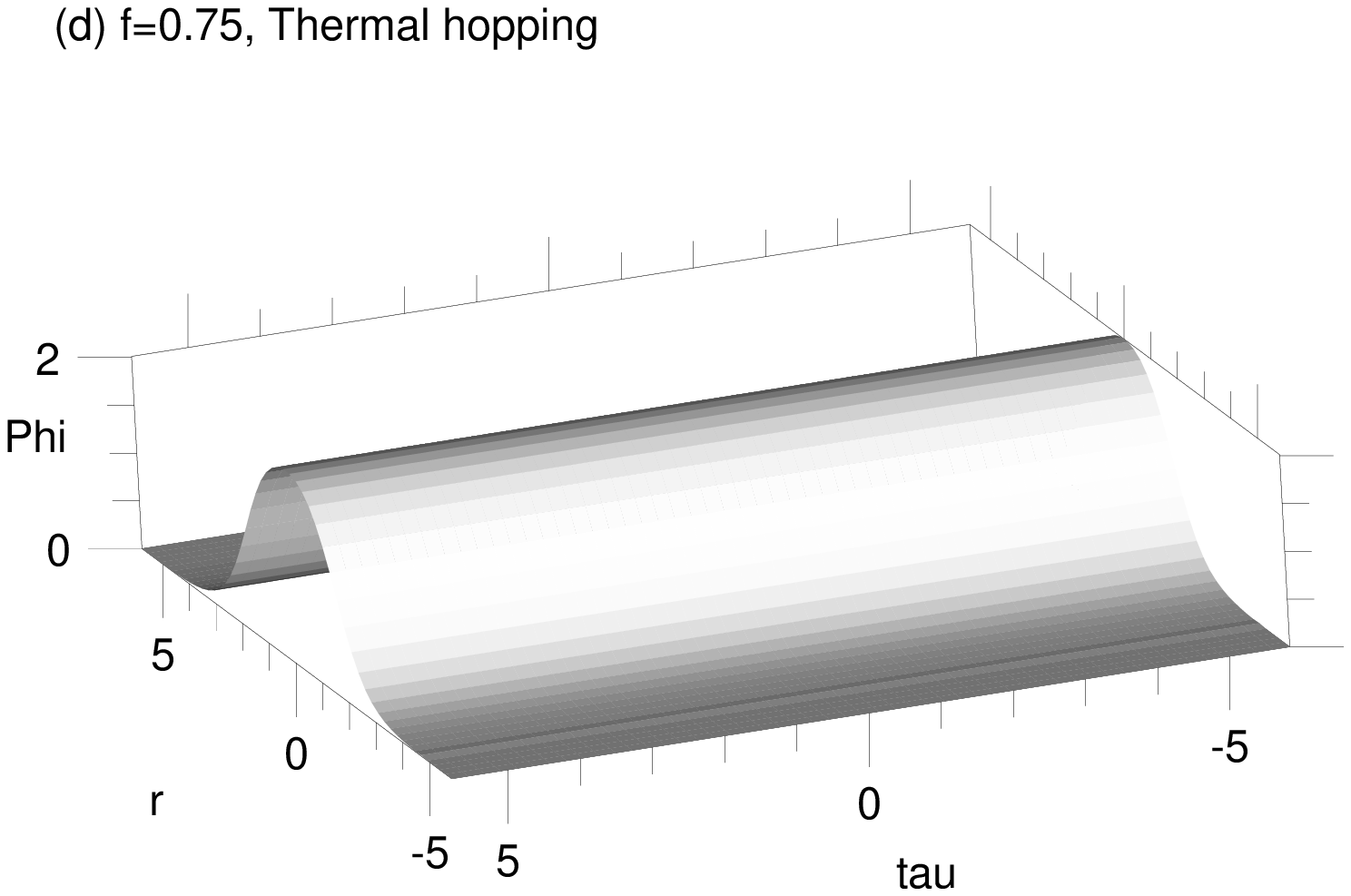} 

\begin{center}
Figure: 9 (d)
\end{center}
\end{figure}
\end{document}